\documentclass{emulateapj}
\RequirePackage{epstopdf}

\slugcomment{Draft of \today}

\shorttitle{WR~112 Paper}
\shortauthors{Lau et al.}

\usepackage{amsmath}
\usepackage{graphicx}
\usepackage[section] {placeins}
\usepackage{hyperref}
\usepackage{color}

\newcommand{\beq}{\begin{equation}}
\newcommand{\eeq}{\end{equation}}

\begin{document}

\title{Stagnant Shells in the vicinity of the Dusty Wolf-Rayet-OB Binary WR~112}

\author{R. M. Lau\altaffilmark{1,2},
M. J. Hankins\altaffilmark{3},
R. Sch{\"o}del\altaffilmark{4},
J. Sanchez-Bermudez\altaffilmark{5},
A. F. J. Moffat\altaffilmark{6},
M. E. Ressler\altaffilmark{1}
}

\altaffiltext{1}{Jet Propulsion Laboratory, California Institute of Technology, 4800 Oak Grove Drive, Pasadena, CA 91109, USA}
\altaffiltext{2}{California Institute of Technology, Pasadena, CA 91125, USA}
\altaffiltext{3}{Astronomy Department, Space Sciences Building, Cornell University, Ithaca, NY 14853-6801, USA}
\altaffiltext{4}{Instituto de Astrof\`{i}sica de Andaluc\`{i}a (CSIC), Glorieta de la Astronom\`{i}a S/N, 18008 Granada, Spain}
\altaffiltext{5}{Max Planck Institut fur Astronomie, K\"{o}nigstuhl 17, 69117 Heidelberg, Germany}
\altaffiltext{6}{D\'{e}partement de physique, Universit\'{e} de Montr\'{e}al, CP 6128, Succ. C.-V., Montr\'{e}al, QC H3C 3J7, and Centre de Recherche en Astrophysique du Qu\'{e}bec, Canada}

\begin{abstract}

We present high spatial resolution mid-infrared images of the nebula around the late-type carbon-rich Wolf-Rayet (WC)-OB binary system WR~112 taken by the recently upgraded VLT spectrometer and imager for the mid-infrared (VISIR) with the PAH1, NeII\_2, and Q3 filters. The observations reveal a morphology resembling a series of arc-like filaments and broken shells. Dust temperatures and masses are derived for each of the identified filamentary structures, which exhibit temperatures ranging from $179_{-6}^{+8}$ K at the exterior W2 filament to $355_{-25}^{+37}$ K in the central 3''. The total dust mass summed over the features is $2.6\pm0.4\times10^{-5}$ $\mathrm{M}_\odot$. A multi-epoch analysis of mid-IR photometry of WR~112 over the past $\sim20$ yr reveals no significant variability in the observed dust temperature and mass. The morphology of the mid-IR dust emission from WR~112 also exhibits no significant expansion from imaging data taken in 2001, 2007, and 2016, which disputes the current interpretation of the nebula as a high expansion velocity ($\sim1200$ km s$^{-1}$) ``pinwheel"-shaped outflow driven by the central WC-OB colliding-wind binary. An upper limit of $\lesssim120$ km s$^{-1}$ is derived for the expansion velocity assuming a distance of $4.15$ kpc. The upper limit on the average total mass-loss rate from the central 3'' of WR~112 is estimated to be $\lesssim8\times10^{-6}$ $\mathrm{M}_\odot$ yr$^{-1}$. We leave its true nature as an open question, but propose that the WR~112 nebula may have formed in the outflow during a previous red or yellow supergiant phase of the central Wolf-Rayet star.


\end{abstract}

\maketitle

\section{Introduction}

Wolf-Rayet (WR) stars are primarily the descendants of massive O-stars that drive powerful winds with terminal speeds $\gtrsim1000$ km/s and high mass-loss rates $\gtrsim10^{-5}$ $\mathrm{M}_\odot\,\mathrm{yr}^{-1}$. Carbon-rich WR (WC) stars, which are identified by broad C emission lines, are unique since many of them are observed to be efficient dust-making factories ($\sim10^{-6}$ $\mathrm{M}_\odot\,\mathrm{yr}^{-1}$ in dust; Gerhz \& Hackwell 1974; Williams et al. 1987; Crowther 2007), despite their hot temperatures ($\mathrm{T}_\mathrm{eff}\sim40,000$ K) and large radiative power output ($\mathrm{L}_*\sim10^5$ $\mathrm{L}_\odot$). A majority of these dusty WC stars are in binary systems with an OB-star companion: strong winds from the WC star collide with weaker winds from the companion and create dense regions in the wake of the companion's orbit that are shielded from the harsh radiation field and allow for dust to condense (e.g. Tuthill et al. 1999). The tell-tale signature of this dust formation process is a remarkable ``pinwheel" that appears to rotate in accordance with the orbital motion of the system (Monnier, Tuthill \& Danchi 1999, 2002; Tuthill et al. 1999, 2006). Dusty WC systems provide a unique laboratory for investigating the mass-loss history of WR binaries since the morphology of the nebula is linked to the orbital parameters of the binary system. These studies can help form a clearer picture of how WR stars evolve in the current paradigm where a majority of massive stars are expected to exchange mass with a close binary companion (Sana et al. 2012).


Observations of these systems have largely been performed using aperture-masking interferometry on near-IR facilities, which probe hot ($\mathrm{T}_\mathrm{dust}\sim1000$ K) inner regions of the pinwheel at high angular resolution (e.g. Tuthill et al. 1999, 2006; Monnier et al. 2007). Near-IR observations, however, are not sensitive to cooler dust that has propagated further from the central system. The late-type WC+OB binary WR~112 (Massey \& Conti 1983; van der Hucht 2001) is one of the few dusty WC systems whose nebula has been detected and resolved in the mid-infrared (8-18 $\mu$m; Marchenko et al. 2002, hereafter referred to as M2002). Others known are WR~48a (Marchenko \& Moffat 2007) and WR~140 (Monnier, Tuthill, \& Danchi 2002; Williams et al. 2009). The well studied WR~140 is the only one of these systems with kinematically measured masses, hosting a 15 M$_\odot$ WC7 star and a 36 M$_\odot$ O5 star (Fahed et al. 2011; Monnier et al. 2011). The resolved mid-IR emission from WR~112, which resembles a series of up to 5 successive broken shells and arc-like filaments, traces cooler dust and provides an opportunity to probe further into the mass-loss history of the central system. Based on mid-IR imaging from Gemini/OSCIR on 2001 May 7, M2002 interpret the WR~112 nebula as a pinwheel produced by the central WC colliding-wind binary with an orbital period of 24.8 yr. 

\begin{figure*}[t!]
	\centerline{\includegraphics[scale=.4]{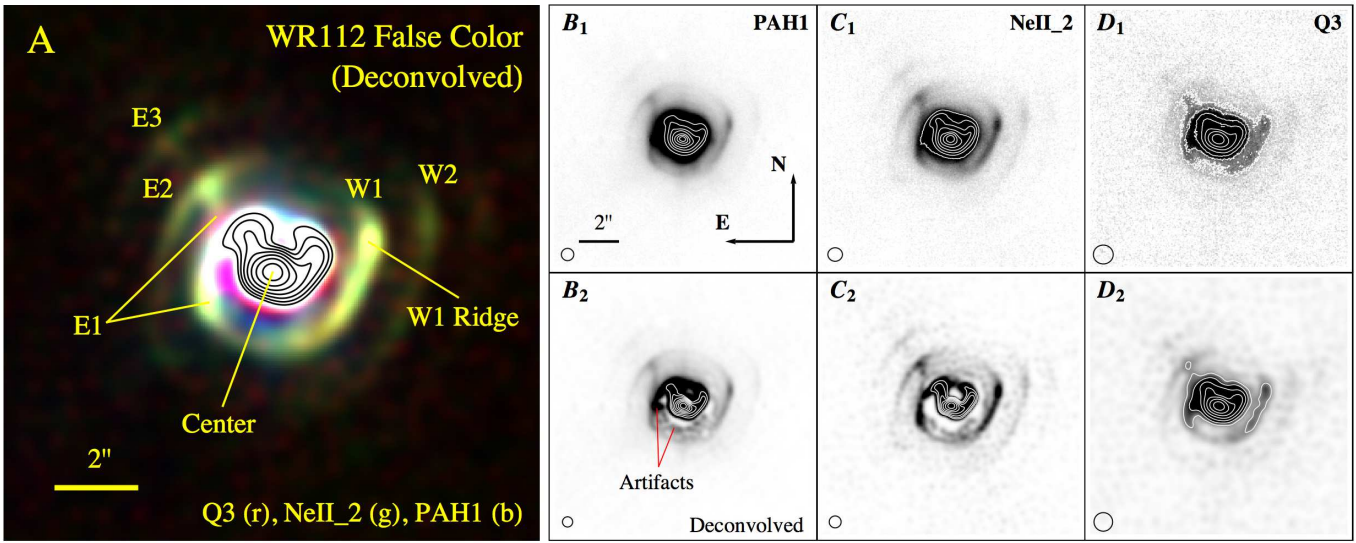}}
	\caption{(A) False-color VLT/VISIR image of WR~112 overlaid with the identified filaments and features. Blue, green, and red colors correspond to emission in the PAH1 ($8.59$ $\mu$m), NeII\_2 ($13.04$ $\mu$m), and Q3 ($19.50$ $\mu$m) filters that have been convolved to a common Gaussian FWHM of $0.5''$. The overlaid black contours correspond to the flux levels at 1, 2, 4, 8, 16, 32, and 64 \% of the peak flux in the NeII\_2 filter. (B$_{1}$ - D$_{1}$) WR~112 observed in the PAH1, NeII\_2, and Q3 filters. The black circles in the lower left represent the FWHM of the standard stars observed on the same night with the same filter (B$_{2}$ - D$_{2}$) Deconvolved images of WR~112 at the identical waveband as the upper row. The overlaid white contours in each image correspond to the flux levels at 1, 2, 4, 8, 16, 32, and 64 \% of the peak flux at each respective filter. Black circles represent the FWHM of the Gaussian PSF used to reconvolve the image. North is up and east is to the left.}
	\label{fig:WR112_Obs}
\end{figure*}

In this letter, we present high spatial resolution imaging of WR~112 using the recently upgraded VLT spectrometer and imager for the mid-infrared (VISIR; Lagage et al. 2004) to investigate the M2002 pinwheel interpretation and the dust properties of the nebula. We perform a multi-epoch analysis from archival space-based IR photometry and spectroscopy of WR~112, and compare our imaging results with that of M2002 and mid-IR observation performed on 2007 May 7 with the Thermal-Region Camera Spectrograph (T-ReCs; De Buizer \& Fisher 2005) on Gemini South (P. ID: GS-2007A-Q-38).

\section{Mid-IR Imaging Observations and Reduction}
\label{Sec:Obs}

Observations of WR~112 (P.ID: 097.D-0707(A); P.I. R. Lau) were performed using VISIR (Lagage et al. 2004) at the Cassegrain focus of UT3. VISIR offers diffraction-limited imaging at high sensitivity in three atmospheric windows: the M-band at 5 $\mu$m, the N-band between $7-14$ $\mu$m, and the Q-band between $17-25$ $\mu$m. The new AQUARIUS detector in VISIR provided a field of view of $38\times38''$ with a plate scale of $0.045''$ per pixel.

Images of WR~112 (R.A. 18:16:33.49, Dec. -18:58:42.3; Cutri et al. 2003) were acquired with the PAH1 ($\lambda_\mathrm{c}=8.59$ $\mu$m, $\Delta \lambda=0.42$), NeII\_2 ($\lambda_\mathrm{c}=13.04$ $\mu$m, $\Delta\lambda=0.22$), and Q3 ($\lambda_\mathrm{c}=19.50$ $\mu$m, $\Delta\lambda=0.40$) filters on 2016 Jul 14, 2016 Aug 9, and 2016 Jul 14, respectively. Chopping and nodding were used to remove the sky and telescope thermal backgrounds. The extent of WR~112 ($<20''$) allowed for an on-chip $20''$-amplitude perpendicular chop-nod observing configuration. The total on-source integration times with the PAH1, NeII\_2, and Q3 filters were 5.6, 5.4, and 23.7 min, respectively. Raw image files were accessed and downloaded from the ESO Science Archive Facility and processed using the Modest Image Analysis and Reduction (MIRA) software written by Terry Herter to reduce and analyze mid-IR imaging data.

The final images of WR~112 were calibrated against mid-IR standard stars obtained within the same night and using the standard-star flux catalog for VISIR imaging filters based on Cohen et al. (1999). A $10\%$ and $20\%$ uncertainty is adopted for the N- and Q-band imaging, respectively, based on the observed long-term variability of the absolute photometric calibration using the VISIR standard stars (Dobrzycka \& Vanzi 2008).

The mean point spread function (PSF) of selected standard stars observed in 2016 was used to deconvolve the PAH1 and NeII\_2 images of WR~112 and by a Richardson-Lucy deconvolution routine with a maximum of $1000$ iterations. Due to the low signal-to-noise ratio and poor stability of the Q3 standards, an Airy function convolved with a Gaussian was instead used as an artificial PSF for deconvolving the Q3 image. The deconvolved PAH1, NeII\_2, and Q3 images were then reconvolved with a Gaussian PSF with FWHM of $0.27''$, $0.36''$, and $0.50''$, respectively.



\section{Results and Analysis}

\subsection{Warm Dust Morphology}

Thermal mid-IR emission from warm dust in WR~112 exhibits a morphology of discontinuous filaments and arcs that are asymmetric about the central flux peak (see Fig.~\ref{fig:WR112_Obs}A) and resembles the shape of a pinwheel as first interpreted by M2002. The morphology of both the bright central $\sim3''$ region and outer filaments is largely consistent across all three wavebands. 

The central 3'' contains $\gtrsim90\%$ of the total emission at each waveband and exhibits a ``U"-shaped morphology. Interestingly, the orientation and shape of the inner mid-IR emission resembles the structure revealed from high-resolution near-IR imaging of WR~112 (FWHM$\sim$20 mas; Monnier et al. 2007) but on larger size scales. A circular arc $\sim1''$ south of the peak is apparent in the deconvolved PAH1- and NeII\_2-band image (Fig.~\ref{fig:WR112_Obs}$\mathrm{B}_2$ and C$_2$), which is likely an artifact from the deconvolution. Additionally, the $\sim0.5''$-sized linear horizontal feature located 1.4'' east of the peak in the PAH1-band image is a detector artifact caused by the position of the bright peak falling on the edge between two detector outputs.

The outer regions of WR~112 consist of a series of broken asymmetric arcs and filaments. These observed structures are referred to as E3, E2, E1, W1, and W2 in Fig.~\ref{fig:WR112_Obs}A). The brightest feature outside of the inner region is located 2.7'' east and slightly north of the central peak and is referred to as the W1 Ridge. The observed properties of the identified structures are summarized in Tab.~\ref{tab:Filaments}.

\subsection{Observed Dust Temperature, Mass, and Luminosity}

The temperature of the emitting warm dust is derived for each morphological component using the fluxes measured from the PAH1- and Q3-band fluxes. It is assumed the emission is optically thin and takes the form of $F_\nu\propto B_\nu(T_d)\nu^{\beta}$, where $B_\nu(T_d)$ is the Planck function at frequency $\nu$ and dust temperature $T_d$, and $\beta$ is the index of the emissivity power-law. A value of -1.5 is adopted for $\beta$, which is consistent with amorphous carbon grains that are believed to compose the nebula (Chiar \& Tielens 2001; M2002). 



The derived temperatures provided in Tab.~\ref{tab:Filaments} are remarkably consistent with the radial temperature profile derived by M2002: $T_0\left(\frac{r}{r_0}\right)^{\gamma}$, where $T_0=320$ K, $r_0=1''$, and $\gamma=-0.4$. For example, the profile predicts a temperature of 222 K at the location of the W1 Ridge, which is consistent with the estimated value of $217^{+12}_{-9}$ K. This decreasing radial temperature profile indicates that the nebula is heated centrally by the WR~112 system. 

Dust masses were derived from the Q3-band flux and dust temperatures of the features as indicated in Eq.~\ref{eq:Mass}:

\beq
M_d=\frac{(4/3)\,a\,\rho_b\,F_\lambda\,d^2}{Q_C(\lambda,a )\,B_\lambda(T_d)},
\label{eq:Mass}
\eeq

\noindent
where $a$ is the adopted grain size, $\rho_b$ is the bulk density of the dust grains, $F_\lambda$ is the measured flux, $d$ is the distance to WR~112, and $Q_C(\lambda,a)$ is the grain emissivity model for amorphous carbon (Zubko et al. 2004). A bulk density of $\rho_b=2.2$ $\mathrm{gm}$ $\mathrm{cm}^{-3}$ (e.g. Draine \& Li 2007) and a grain size of $a=0.5$ $\mu$m (M2002) is assumed for the emitting amorphous carbon. The dust mass estimated for each feature is provided in Tab.~\ref{tab:Filaments}. The total dust mass summed over all of the features is $2.6\pm0.4\times10^{-5}$ $\mathrm{M}_\odot$, which again shows a remarkable agreement with the dust mass estimated from M2002 ($2.8\times10^{-5}$  $\mathrm{M}_\odot$). 



\begin{figure}[t!]
	\centerline{\includegraphics[scale=.5]{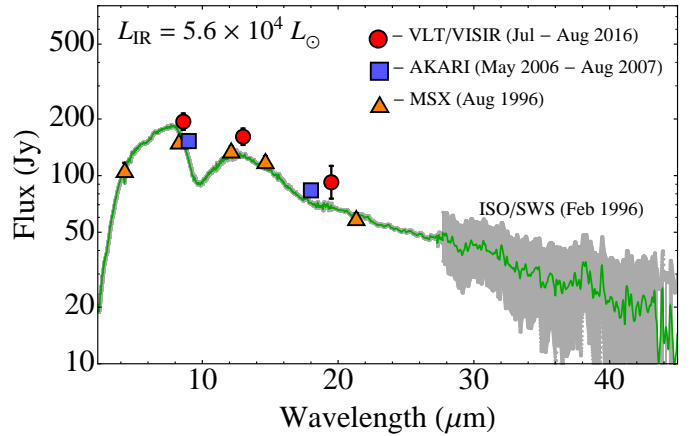}}
	\caption{Multi-epoch IR photometry and spectroscopy of WR~112. The gray lines around the ISO/SWS spectrum correspond to the $1-\sigma$ flux uncertainty. Photometry with no visible error bars indicates that the errors are smaller than the size of the plot marker.}
	\label{fig:WR112_SED}
\end{figure}

In Fig.~\ref{fig:WR112_SED}, archival spectroscopic and photometric IR observations of WR~112 from various platforms over the past $\sim20$ yr are presented. The IR space-based data shown in Fig.~\ref{fig:WR112_SED} were acquired from the Infrared Space Observatory (ISO; Kessler et al. 1996) Short Wavelength Spectrometer (SWS; de Graauw et al. 1996), SPIRIT III on the Midcourse Space Experiment (MSX; Mill et al. 1994), and the Infrared Camera (IRC) All-Sky Survey (Ishihara et al. 2010) from AKARI (Murakami et al. 2007). The photometry of WR~112 from MSX and AKARI received the highest quality indicator. A total observed IR luminosity ($2.4-45.4$ $\mu$m) of L$_\mathrm{IR}=5.6\times10^{4}$ L$_\odot$ was determined from the ISO/SWS spectrum assuming a distance of 4.15 kpc. The IR luminosity is consistent with re-radiating a significant fraction of WR~112's stellar luminosity of $L_*=9\times10^{4}$ L$_\odot$, which is derived from its absolute $V$-band magnitude ($M_V=-4.62$, van der Hucht 2001) and bolometric correction of BC$_\mathrm{V}=3.0$ (Smith, Meynet, Mermilliod 1994).

Despite observations of variable near-IR (Williams et al. 2015) and radio (Monnier et al. 2002; Yam et al. 2015) emission, the mid-IR data do not indicate significant variations in the dust mass or temperature over time. The orbital properties of the central colliding-wind binary system, which are presumably linked to the variable near-IR and radio emission, may therefore not be associated with the production of the observed dust in the nebula. Importantly, the assumed 24.8 yr period of the WR~112 binary (M2002) is well-sampled by the IR observations taken at 1996 (ISO and MSX), 2001 (Gemini), 2006-07 (AKARI), and 2016 (VLT).

\section{Discussion and Future Work}

\begin{figure*}[t!]
	\centerline{\includegraphics[scale=0.67]{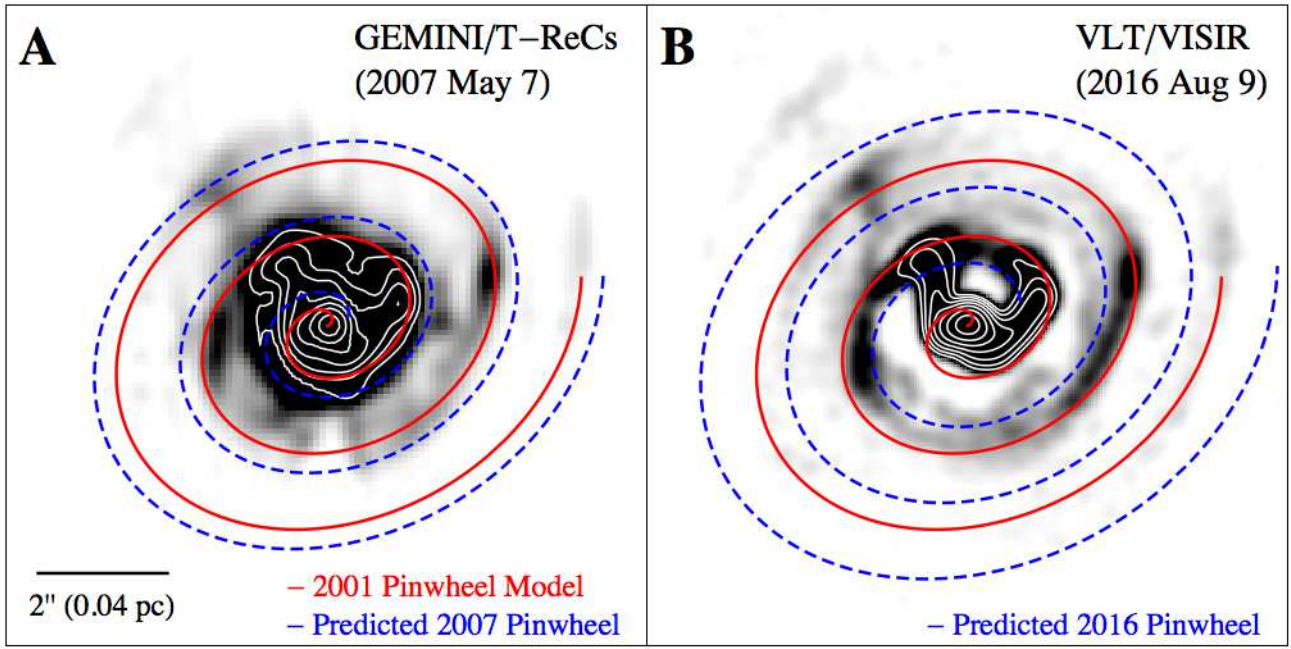}}
	\caption{(A) Deconvolved Gemini/T-ReCs Si5 and (B) VLT/VISIR NeII\_2 images of WR~112 overlaid with the high eccentricity ``Pinwheel" model fit by M2002 to mid-IR images of WR~112 taken in 2001 (red solid line). Blue dashed lines represent the predicted appearance of the nebula according to the pinwheel model on the dates the images in (A) and (B) were taken. The white contours in each image correspond to the flux levels at 1, 2, 4, 8, 16, 32, and 64 \% of the peak flux. North is up and east is to the left.}
	\label{fig:WR112_Spiral}
\end{figure*}

\subsection{Comparison to the ``Pinwheel" Model}

The current interpretation of the WR~112 nebula is that of a broken ``Pinwheel" produced by a central colliding-wind late-type WC9+OB binary system (M2002; Monnier et al. 2002, 2007) due to the regularly spaced features and their similar appearance. Under this interpretation, one can estimate the orbital period of the central system and predict the appearance of the nebula assuming an expansion velocity. Our mid-IR imaging data provides a sufficient temporal baseline against previous resolved mid-IR images (M2002) to test the validity of the pinwheel model. Additionally, we include an intermediary mid-IR imaging epoch of WR~112 taken by Gemini/T-ReCs\footnote{The Gemini/T-ReCs image was deconvolved using the same routine as the VISIR images (Sec.~\ref{Sec:Obs}), except a Gaussian with a FWHM of $0.45''$ was adopted for the PSF.} on 2007 May 7 (P. ID: GS-2007A-Q-38) with the Si5 filter ($\lambda_\mathrm{eff}=11.6$ $\mu$m) in order to identify morphological variations on shorter timescales. 

We adopt the high-eccentricity colliding-wind binary/pinwheel model fit by M2002 to their 2001 mid-IR images of WR~112 and compare their results against the 2007 and 2016 images (Fig.~\ref{fig:WR112_Spiral}A and B). The model parameters assume an expansion velocity of 1200 km $\mathrm{s}^{-1}$ and the following orbital parameters for the central the binary system: $i=35^\circ$, $\Omega=130^\circ$, $\omega=165^\circ$, $e=0.40$, and $P = 23.5$ yr. Predicted 2007 and 2016 pinwheel models overlaid in Fig.~\ref{fig:WR112_Spiral}A and B (blue dashed lines) are derived by propagating the 2001 model assuming a velocity of 1200 km $\mathrm{s}^{-1}$. By comparing the pinwheel models and imaging data, we draw the following conclusions:

1. The mid-IR nebula is not propagating in accordance with the colliding-wind binary/pinwheel model proposed by M2002.

2. There is no apparent expansion of the nebula in the 6, 9, and 15 yr intervals between the 2001, 2007, and 2016 observations.



The``stagnant" morphology implies an upper limit on the expansion velocity of $\lesssim120$ km $\mathrm{s}^{-1}$ between 2007 and 2016 images assuming a $\sim1$ VISIR pixel (0.045'') centroid alignment uncertainty, a distance of 4.15 kpc, and an inclination of $i=35^\circ$. The low velocity is unlikely due to geometric projection effects since a near edge-on inclination of $i\sim85^\circ$ would be required to infer a velocity consistent with the predicted 1200 km s$^{-1}$ WR outflow. Such a high inclination is inconsistent with the near face-on and circular appearance of the nebula.

The upper limit derived for the expansion velocity is almost an order of magnitude less than the predicted and observed velocities of dusty outflows produced in colliding-wind WR binaries like WR~140 (Williams et al. 2009) and WR~104 (Tuthill et al. 1999). Additionally, the mid-IR fluxes and inferred dust properties of WR~112 (Tab.~\ref{tab:FluxProp}) do not exhibit significant variability, whereas the orbitally-modulated dust production in WR~140 shows an order of magnitude variation in 8.75 and 12.5 $\mu$m emission (Williams et al. 2009). It is therefore unlikely that the nebula originated from the high velocity outflows of the central WR binary. An alternative dust production scenario is thus required to explain the presence of the nebula.

\subsection{On the Nature of the WR~112 Nebula: Mass-loss from a Previous Evolutionary Phase?}

Here, we discuss a possible scenario that can account for stagnant or slowly expanding shells in the vicinity of the WR~112 system. It is important to note that previous studies have spectroscopically verified the presence of a WC9 star at the center of WR~112 (Massey \& Conti 1983; Figer, McLean, \& Najarro 1997). Additionally, a binary companion is inferred from variable non-thermal radio emission that is believed to originate from a central wind-collision zone (Chapman et al. 1999; Monnier et al. 2002). The unresolved radio counterpart of WR~112 exhibits a peculiar motion of $-100\pm34$ km $\mathrm{s}^{-1}$ in Galactic longitude (Yam et al. 2015), which is not consistent with the position angle of the shells nor the central ``U"-shaped region if they were instead bow shocks from the interstellar medium. 

Larger-sized nebulae ($\gtrsim1$ pc) surrounding WR stars are commonly interpreted as ejecta from a previous phase of high mass-loss. The ejecta can originate from the slow and dense winds of the star during its luminous blue variable (LBV; $v\sim100$ km $\mathrm{s}^{-1}$) or red supergiant (RSG; $v\sim30$ km $\mathrm{s}^{-1}$) phase, later interacting with the high velocity winds and hard radiation field from the subsequent WR phase (e.g. Garcia-Segura \& Mac Low 1995; Freyer et al. 2006; Toal\'{a} et al. 2015). Another potential origin for the WR~112 nebula is episodic mass-loss during a short-lived post-RSG yellow supergiant (YSG) phase, where the star undergoes a blueward evolution to the WR phase (e.g. de Jager 1998; Smith et al. 2004; Humphreys et al. 2013; Gordon et al. 2016).




We attempt to distinguish amongst an LBV, YSG, and RSG phase for the origin of the WR~112 nebula by estimating upper limits on the average mass-loss rate. Assuming a gas-to-dust mass ratio of $100$ and an expansion velocity $<120$ km $\mathrm{s}^{-1}$, the average mass-loss rate from the central 3'' (0.06 pc) region is $\dot{M}<8\times10^{-6}$ $\mathrm{M}_\odot$ yr$^{-1}$. This upper limit is several orders of magnitude lower than the observed mass-loss rates from dust-forming LBVs during giant eruptions ($\gtrsim10^{-3}$ $\mathrm{M}_\odot$ yr$^{-1}$; Kochanek 2011; Smith 2014);
however, the mass-loss rate and total dust mass in the nebula is consistent with values derived from circumstellar material surrounding RSGs and YSGs ($\dot{M}\sim10^{-5}-10^{-4}$ $\mathrm{M}_\odot$ yr$^{-1}$, M$_\mathrm{d}\sim10^{-5}-10^{-4}$ M$_\odot$; Gordon et al. 2016). 
Therefore, if the nebula formed prior to the WR phase it most likely originated from an RSG or post-RSG/YSG outflow, but we cannot decisively rule out an LBV origin. 
We note that these evolved phases of massive stars typically exhibit an oxygen-rich chemistry, which conflicts with the inferred carbon-rich composition of the nebula (Chiar \& Tielens 2001; M2002). However, the presence of amorphous carbon was determined from spatially unresolved mid-IR spectroscopy dominated by emission from the central $\sim3''$ of WR~112. Identifying chemical differentiation between the central and extended regions of the nebula would aid in testing the RSG/YSG hypotheses.


The YSG interpretation is interesting from an evolutionary standpoint because the proximity of the nebula to the central system and its low expansion velocity imply that the WR star has newly transitioned. However, the carbon-rich photosphere of WR~112 indicates it is in the most evolved stage of the WR sequence (e.g. Crowther 2007). An estimate of the dynamical age of the W1 filament assuming an YSG-like outflow velocity of $50$ km s$^{-1}$ implies that WR~112 may have transitioned within the past $\sim1000$ yr, close to the observationally estimated lifetime of YSGs ($\sim3000$ yr; Drout et al. 2009) but two orders of magnitude less than the predicted WR star lifetime ($\sim10^5$ yr; Meynet \& Maeder 2005). One possible explanation for the rapid evolution of WR~112 may be mass-transfer via interactions with a close binary companion (e.g. Smith et al. 2011a). Two notable examples of observed mass transfer likely leading to a stripped-envelope WR star are NaSt1 (Mauerhan et al. 2015) and RY Scuti (Smith et al. 2011b). WR~112 may therefore be another such example of binary interaction influencing the evolution of massive stars. 


Ultimately, we leave the true nature of WR~112 as an open question since we are only able to conclude that the stagnant nebula is inconsistent with dusty WC outflows. 



\subsection{Future Work}


High spatial resolution mid-IR spectroscopy of WR~112 would provide crucial information on the dust composition and chemical abundances throughout the nebula to verify the origin of their formation and the conditions of grain growth. Given the sensitivity limitations of current ground-based platforms due to Earth's atmosphere, we require the combined sensitivity and spatial resolution that will be achievable with the Mid-Infrared Instrument (MIRI; e.g. Wells et al. 2015) on the \textit{James Webb Space Telescope} (JWST), which is expected to launch Oct 2018. JWST will be the ideal platform for investigating the origin of the compact warm nebulae around dusty WC systems and deciphering the mass loss history and evolution of their central engines. 




\emph{Acknowledgments}. 

This work is based on observations made with the VISIR instrument on the ESO VLT telescope (program ID. 097.D-0707A) and observations obtained at the Gemini Observatory (P. ID: GS-2007A-Q-38), which is operated by the Association of Universities for Research in Astronomy, Inc., under a cooperative agreement with the NSF on behalf of the Gemini partnership: the National Science Foundation (United States), the National Research Council (Canada), CONICYT (Chile), Ministerio de Ciencia, Tecnolog\'{i}a e Innovaci\'{o}n Productiva (Argentina), and Minist\'{e}rio da Ci\^{e}ncia, Tecnologia e Inova\c{c}\~{a}o (Brazil). This work was partially carried out at the Jet Propulsion Laboratory, California Institute of Technology, under a contract with the National Aeronautics and Space Administration. M.J.H. acknowledges support from the National Science Foundation Graduate Research Fellowship under Grant No. DGE-1144153. J.S.B acknowledges that this work was partly supported by OPTICON, which is sponsored by the European Commission's FP7 Capacities programme (Grant number 312430). AFJM is grateful for financial aid to NSERC (Canada) and FQRNT (Quebec). R.S. acknowledges funding from the European Research Council under the European Union's Seventh Framework Programme (FP7/2007-2013) / ERC grant agreement n$^\mathrm{o}$ [614922].

R.L. would like to thank Sergey Marchenko for valuable feedback and comments, and Jim De Buizer and James Radomski for the helpful insight on T-ReCs imaging data. R.L. also thanks Astrid Lamberts for the enlightening conversations on colliding wind binaries. Lastly, we thank the anonymous referee for the insightful suggestions and comments.

\newpage

\clearpage

\section{Tables}

\begin{deluxetable}{cccccccc}
\tablecaption{Observed fluxes and dust properties throughout the WR~112 nebula}
\tablewidth{0pt}
\tablehead{ Region & $l$ ('') & $d$ ('') & $\mathrm{F}_\mathrm{PAH1}$ & $\mathrm{F}_\mathrm{NeII\_2}$ & $\mathrm{F}_\mathrm{Q3}$ & $\mathrm{T}_\mathrm{d}$ (K) & $\mathrm{M}_\mathrm{d}$ $(10^{-6}\,\mathrm{M}_\odot)$  }

\startdata
W1& 3.7 &2.5&1.16&2.44&2.31&$216_{-9}^{+12}$&$2.5_{-0.4}^{+0.4}$ \\ 
W2& 4.5  &4.2&0.1&0.49&0.48&$179_{-6}^{+8}$&$1.1_{-0.2}^{+0.2}$ \\ 
W1 Ridge& 1.0 &2.5&0.3&0.68&0.58&$217_{-9}^{+12}$&$0.6_{-0.1}^{+0.1}$ \\ 
E1& 3.6&1.8&1.93&3.17&2.82&$233_{-10}^{+14}$&$2.4_{-0.4}^{+0.4}$ \\ 
E2& 3.6&2.7&0.36&0.87&0.85&$207_{-8}^{+11}$&$1.1_{-0.2}^{+0.2}$ \\ 
E3*& 2.7 &4.2&0.12&0.31&0.19& -- & -- \\ 
Center & 3 &$<1.5$&179.3&140.&71.37&$355_{-25}^{+37}$&$18.6_{-3.8}^{+3.6}$ \\ 

\enddata

\tablecomments{Fluxes given are provided in Jy. The approximate length, $l$, and distance from the central peak, $d$, of each feature is given in arc seconds. \\ * - Dust temperatures and masses were not derived for the E3 filament due to the low signal-to-noise ratio of the detection in the Q3 image.}
	\label{tab:Filaments}
\end{deluxetable}

\begin{deluxetable}{cccccc}
\tablecaption{Multi-Epoch IR Photometry and Derived Dust Properties of WR~112}
\tablewidth{0pt}
\tablehead{ Date & Observatory & Band (Wavelength) & Flux (Jy)  & $\mathrm{T}_\mathrm{d}$ (K) & $\mathrm{M}_\mathrm{d}$ $(10^{-5}\,\mathrm{M}_\odot)$  }

\startdata
Aug 1996 & MSX/SPIRIT III & B1 $(4.29\,\mu\mathrm{m})$ & $107.2\pm9.3$ & $335_{-7}^{+7}$ & $2.2_{-0.1}^{+0.1}$  \\
 		& 		 & A $(8.28\,\mu\mathrm{m})$& $152\pm6.2$ & &  \\
 		& 		 & C $(12.13\,\mu\mathrm{m})$& $136\pm6.8$ & &  \\
		& 		 & D $(14.65\,\mu\mathrm{m})$& $120\pm7.3$ & &  \\
		& 		 & E $(21.34\,\mu\mathrm{m})$& $59\pm3.6$ & &  \\
May 2006 - Aug 2007 & AKARI/IRC & S9W $(9\,\mu\mathrm{m})$ & $152.6\pm0.86$&  $329_{-1}^{+1}$ & $2.2_{-0.02}^{+0.02}$  \\
 		& 		 & L18W $(18\,\mu\mathrm{m})$& $83.5\pm0.15$& &  \\
Jul 2016 & VLT/VISIR & PAH1 $(8.59\,\mu\mathrm{m})$ & $194\pm19$& $331_{-21}^{+31}$ &  $2.8_{-0.6}^{+0.5}$ \\
 Aug 2016& 		 & NeII\_2 $(13.04\,\mu\mathrm{m})$& $161\pm16$& &  \\
 Jul 2016 & 		 & Q3 $(19.50\,\mu\mathrm{m})$& $92\pm18$ & &  \\

\enddata

\tablecomments{Dust temperatures, T$_\mathrm{d}$, were derived from the A/E, S9W/L18W, and PAH1/Q3 bands in for the observations provided by MSX, AKARI, and VLT, respectively. Dust masses, M$_\mathrm{d}$, were then derived from Eq.~\ref{eq:Mass} assuming identical dust properties that were used to provide the values in Tab.~\ref{tab:Filaments}}
	\label{tab:FluxProp}
\end{deluxetable}

\end{document}